%% file: bare_jrnl_new_sample4.tex
\documentclass[lettersize,journal]{IEEEtran}
\usepackage{amsmath,amsfonts}
\usepackage{algorithmic}
\usepackage{algorithm}
\usepackage{array}
\usepackage{lipsum}
\usepackage[caption=false,font=normalsize,labelfont=sf,textfont=sf]{subfig}
\usepackage{textcomp}
\usepackage{stfloats}
\usepackage{url}
\usepackage{verbatim}
\usepackage{graphicx}
\usepackage{cite}
\usepackage{hyperref}
\usepackage{multirow}
\usepackage{multirow}
 \usepackage{multirow}
\usepackage{array}
\usepackage{titlesec}
\usepackage{hyperref}
\usepackage[table,xcdraw]{xcolor}
\usepackage{multirow}
\usepackage[table,xcdraw]{xcolor}
\usepackage{geometry}
\usepackage{graphicx} 
\usepackage{array}    
\usepackage{booktabs} 
\usepackage[table,xcdraw]{xcolor}
\usepackage{colortbl}
\usepackage{float}
\usepackage{soul}
\usepackage{titlesec}
\usepackage{balance}
\usepackage{amsmath}
\usepackage{multirow}
\usepackage{graphicx}
\usepackage{array}
\usepackage{booktabs}
\usepackage{tabularx}
\usepackage{multirow}
\usepackage{colortbl}
\usepackage{array}
\usepackage{colortbl}
\usepackage{array}
\usepackage{xcolor} 
\usepackage{booktabs} 
\arrayrulecolor{black}
\usepackage{array}
\usepackage{xcolor}
\usepackage{colortbl}
\usepackage{tabularx} 
\usepackage{graphicx} 
\usepackage[table,xcdraw]{xcolor}
\usepackage[table,xcdraw]{xcolor} 

\usepackage{pdflscape} 
\usepackage{graphicx} 
\usepackage{array} 
\usepackage{longtable} 
\usepackage{lscape}
\usepackage{graphicx} 
\usepackage{array} 
\usepackage{caption}
\usepackage{afterpage}
\usepackage{caption}
\usepackage{amsmath}
\usepackage{multirow}
\usepackage{array}
\usepackage{booktabs}
\usepackage{tabularx}
\usepackage[table]{xcolor}
\usepackage{etoolbox}
\usepackage{url}
\usepackage{microtype}
\usepackage{amsmath} 
\usepackage[abbreviations]{glossaries-extra}
\glsdisablehyper

\usepackage[table,xcdraw]{xcolor}
\definecolor{custompurple}{RGB}{128,0,128} 

\definecolor{customblue}{RGB}{173, 216, 230} 
\definecolor{nonisolated}{RGB}{230, 255, 230}  
\definecolor{isolated}{RGB}{230, 240, 255}     
\definecolor{noniso1}{RGB}{235, 255, 235}
\definecolor{noniso2}{RGB}{215, 245, 215}
\definecolor{noniso3}{RGB}{195, 235, 195}
\definecolor{noniso4}{RGB}{175, 225, 175}
\definecolor{noniso5}{RGB}{155, 215, 155}
\definecolor{noniso6}{RGB}{135, 205, 135}
\definecolor{noniso7}{RGB}{115, 195, 115}
\definecolor{noniso8}{RGB}{95, 185, 95}
\definecolor{noniso9}{RGB}{75, 175, 75}
\definecolor{noniso10}{RGB}{55, 165, 55}
\definecolor{noniso11}{RGB}{35, 155, 35}
\definecolor{noniso12}{RGB}{15, 145, 15}

\definecolor{iso1}{RGB}{235, 240, 255}
\definecolor{iso2}{RGB}{215, 225, 250}
\definecolor{iso3}{RGB}{195, 210, 245}
\definecolor{iso4}{RGB}{175, 195, 240}
\definecolor{iso5}{RGB}{155, 180, 235}
\definecolor{iso6}{RGB}{135, 165, 230}

\begin{document}
\bstctlcite{IEEEexample:BSTcontrol}

\raggedbottom
\title{Adaptive Gradient Descent MPPT Algorithm With Complexity-Aware Benchmarking for Low-Power PV Systems}


\author{Kimia Ahmadi,~\IEEEmembership{Student Member,~IEEE}, Wouter A. Serdijn,~\IEEEmembership{Fellow,~IEEE}


\thanks{Kimia Ahmadi and Wouter A. Serdijn are with the Department of Microelectronics, Delft University of Technology, The Netherlands.}
\thanks{This publication is part of the project Dutch Brain Interface Initiative (DBI\textsuperscript{2}) with project number 024.005.022 of the research programme Gravitation, which is financed by the Dutch Ministry of Education, Culture and Science (OCW) via the Dutch Research Council (NWO).} 
}



\maketitle  

\begin{abstract}

This paper proposes a computationally efficient, real-time maximum power point tracking (MPPT) algorithm tailored for low-power photovoltaic (PV) systems operating under fast-changing irradiance and partial shading conditions (PSC). The proposed method augments the classical perturb and observe (P\&O) algorithm with an adaptive gradient descent mechanism that dynamically scales the perturbation step size based on the instantaneous power–voltage slope, thereby minimizing tracking time and steady-state oscillations. An optional initialization routine enhances global MPP (GMPP) tracking under PSC. Extensive simulations, including experimental-derived irradiance data from freely moving rodent subjects relevant for the targeted application, and tests across varying converter topologies and temperatures, demonstrate its robust, topology-independent performance. The proposed algorithm achieves 99.94\% MPPT efficiency under standard test conditions (STC), 99.21\% when applied on experimental data, and $>$99.6\% for the tested temperature profiles. Under PSC, the initialization routine improves tracking efficiency by up to 7.8\%.
A normalized gate-level complexity analysis and a unified figure-of-merit (FoM) incorporating efficiency, tracking time, and computational cost demonstrate that the proposed algorithm outperforms 35 state-of-the-art P\&O-based MPPT algorithms. These results underscore its suitability for integration in low-power power management integrated circuits (PMICs) operating under dynamic and resource-constrained conditions.

\end{abstract}


\section{Introduction} \label{sec:intro}

Optical wireless power transfer (OWPT) enables efficient, directed wireless power transfer (WPT) where wired or conventional links are impractical, making it suitable for mobile applications such as biomedical wearables and internet of things (IoT) nodes. However, receiver motion and environmental variability lead to fluctuating incident power, complicating reliable energy harvesting. Robust maximum power point tracking (MPPT) is thus essential to maximize extraction from photovoltaic (PV) receivers under dynamic conditions \cite{Ahmadi2025,DiPatrizioStanchieri2023}.

The traditional MPPT algorithms, such as perturb and observe (P\&O) and incremental conductance (IC), provide simplicity but suffer under dynamic conditions. Intelligent and hybrid techniques, including artificial neural networks (ANN), and metaheuristic methods such as particle swarm optimization (PSO), provide high accuracy at a high computational cost or offline training and hence are not suitable for real-time embedded systems \cite{Khairi2023,T2024,VietAnh2024,Zhao2023}.


Partial shade conditions (PSC) introduce multiple local maxima into the power–voltage (P-V) curve, making MPPT challenging. Advanced techniques like gray wolf optimization (GWO), genetic algorithms (GA), etc., enhance global maximum power point (GMPP) tracking but entail significant computation burden or off-line training \cite{Belmadani2024, Kumari2024, Etezadinejad2022}. Model-based and off-line methods provide alternatives but are not viable for unsupervised applications. Temperature changes further affect MPPT by shifting the maximum power point (MPP). Some recent
research \cite{Chellakhi2021,Jabbar2023} reported degraded tracking under thermal alterations. Intelligent techniques such as long-short-term memory (LSTM)-based MPPT \cite{Tang2024} provide improved thermal robustness but come at higher computational cost. This work analyzes the impact of temperature variations on a new adaptive MPPT algorithm applicable to dynamic OWPT scenarios.




Motivated by the Dutch Brain Interface Initiative (DBI\textsuperscript{2}) project \cite{DBI2_2025}, focused on neural recording with a head-mounted optical receiver for freely moving rodents, this paper addresses the challenges of rapid movements causing irradiance fluctuations. The key contributions include: (1) an adaptive gradient descent–enhanced P\&O algorithm for low-power dynamic OWPT systems that scales the perturbation step with the real-time power–voltage gradient to suppress steady-state oscillations and accelerate convergence; (2) a light-weight initialization routine enhancing global MPPT under PSC; (3) an input-independent, converter-agnostic framework for real-time embedded applications; (4) a normalized computational load evaluation based on gate-level ASIC operation costs; and (5) an integrated figure of merit (FoM) balancing efficiency, tracking time, and complexity for equitable benchmarking across various algorithms irrespective of the system's power level. Collectively, these provide a rigorous foundation for hardware-aware, efficient, and scalable MPPT across dynamic OWPT.


The rest of the paper is organized as follows: Section II describes the designed photovoltaic system. Section III presents the proposed adaptive gradient descent–based MPPT algorithm and initialization routine. Section IV presents simulation results under varying irradiance, temperature, converter types, and PSC, along with computational complexity benchmarking. Section V establishes a unified FoM for comparison of algorithms under fair conditions. Finally, section VI concludes with observations and future directions.

\section{Photovoltaic System Overview}



A detailed understanding of the PV module and the power conditioning stage is important for the design of an effective MPPT for OWPT systems. This section describes these two core components to support the subsequent algorithm development and analysis.

\subsection{PV Module Modeling}




We use the single-diode PV model in Fig.~\ref{fig:single diode PV electrical model}, with $R_s$ and $R_{sh}$ denoting series and shunt resistances, defines the output current as in (\ref{eq:1}):\cite{Villalva2009}:\vspace{-1.6em}

\begin{figure}[H]

\centering
\includegraphics[width=2in]{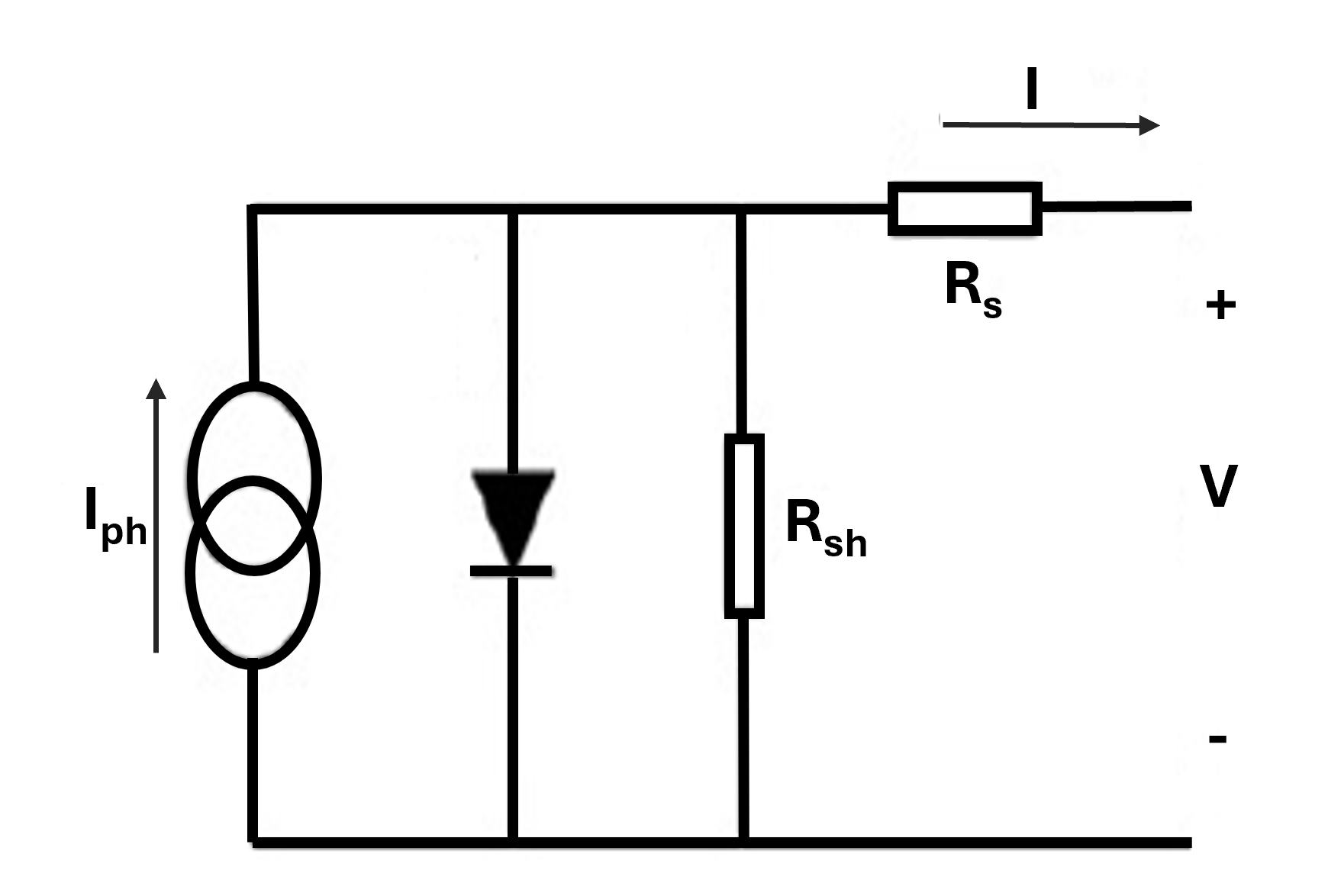}
\caption{Single-diode equivalent circuit of a PV cell.}
\label{fig:single diode PV electrical model}
\end{figure}

\vspace{-10pt} 

\begin{equation}
I = I_{\mathrm{ph}} - I_s \left[\exp\left(\frac{V + IR_s}{\alpha V_t}\right) - 1\right] - \frac{V + IR_s}{R_{sh}}
\label{eq:1}
\end{equation}

 In (\ref{eq:1}), the $V$ is the terminal voltage, $I_{\mathrm{ph}}$ is the photocurrent, $I_s$ is the diode saturation current, $\alpha$ is the diode ideality factor, and $V_t=$ $\frac{k T}{q_e}$ is the thermal voltage, with $k$ being Boltzmann's constant, $T$ the absolute temperature, and $q_e$ the electron charge.


\begin{equation}
I_{\mathrm{ph}} = \left(I_{\mathrm{sc,STC}} + k_i(T - T_{\mathrm{STC}})\right) \cdot \frac{G}{G_{\mathrm{STC}}}
\label{eq:2}
\end{equation}

\begin{equation}
I_s = \frac{I_{\mathrm{sc,STC}} + k_i(T - T_{\mathrm{STC}})}{\exp\left(\frac{V_{\mathrm{oc}} + k_v(T - T_{\mathrm{STC}})}{aV_t}\right) - 1}
\end{equation}


In (\ref{eq:2}), $I_{\mathrm{sc,STC}}$ and $V_{\mathrm{oc}}$ are the short-circuit current and open-circuit voltage at STC;   ($\mathrm{T}_{\mathrm{STC}}=25^{\circ} \mathrm{C}$ and $\mathrm{G}_{\mathrm{STC}}=1000 \mathrm{~W} / \mathrm{m}^2$), $k_i$ and $k_v$ are temperature coefficients, and $G$ is the incident irradiance.

\subsection{DC-DC Boost Converter}

\begin{figure}[H]
\vspace{-10pt} 
\centering
\includegraphics[width=3.5in]{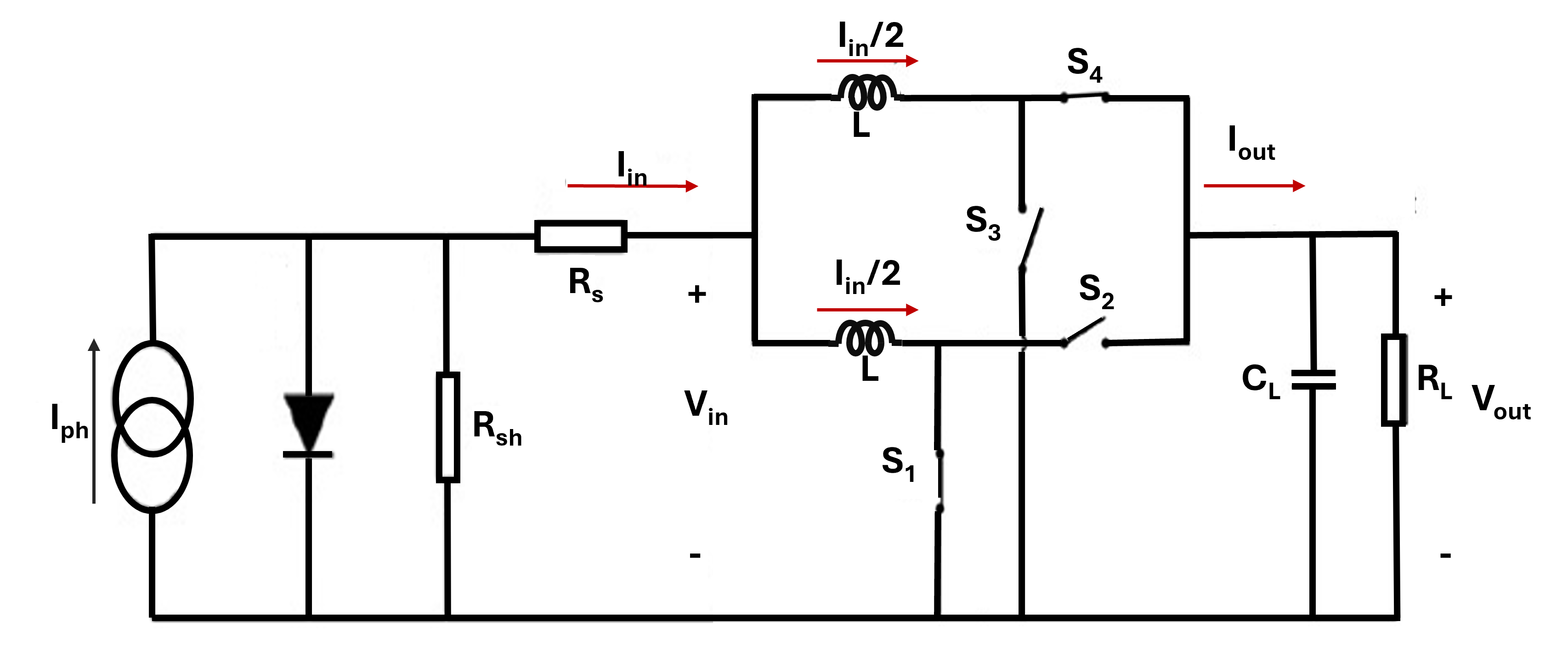}
\caption{2-phase interleaved boost converter.}
\label{fig:2 phase interleaved}
\end{figure}

The DC–DC converter ensures PV-load impedance matching by regulating energy transfer through the duty cycle. While various topologies, such as boost, buck–boost, and multilevel converters, offer trade-offs in complexity and efficiency \cite{Forouzesh2016,Kahani2023,Das2010}, often suffer from high input current ripple when operated at high voltage gains \cite{Mamarelis2014,Santra2023}. The 2-phase interleaved boost converter (Fig.~\ref{fig:2 phase interleaved}) addresses these problems by sharing the input current between the two branches, reducing critical inductor current ripple and boosting conversion efficiency at moderate complexity \cite{Roy2017,Parizi2024,Khorasani2022}. As proof of principle, we use a 2-phase interleaved converter. However, the proposed algorithm also works for other converter types.

The 2-phase interleaved converter's ideal voltage gain remains identical to that of a conventional boost converter. The switches operate in complementary pairs, and the conduction loss is dominated by the equivalent series resistances (ESR) of the inductors and the on-resistances of the switches. Losses due to gate charging are proportional to device capacitance and switching frequency. The robustness of the proposed adaptive MPPT algorithm, under dynamic conditions, is verified through comprehensive simulations in the following sections.

\section{proposed MPPT Algorithm}

This section presents a novel adaptive gradient descent–based P\&O MPPT algorithm for low-power OWPT systems, featuring adaptive step size control and an optional initialization for robust performance at PSC.

\subsection{Overview of conventional P\&O Algorithm} \label{subsec:overview_PO}



\begin{figure}[H]
\vspace{-8pt}
\centering
\includegraphics[width=3in]{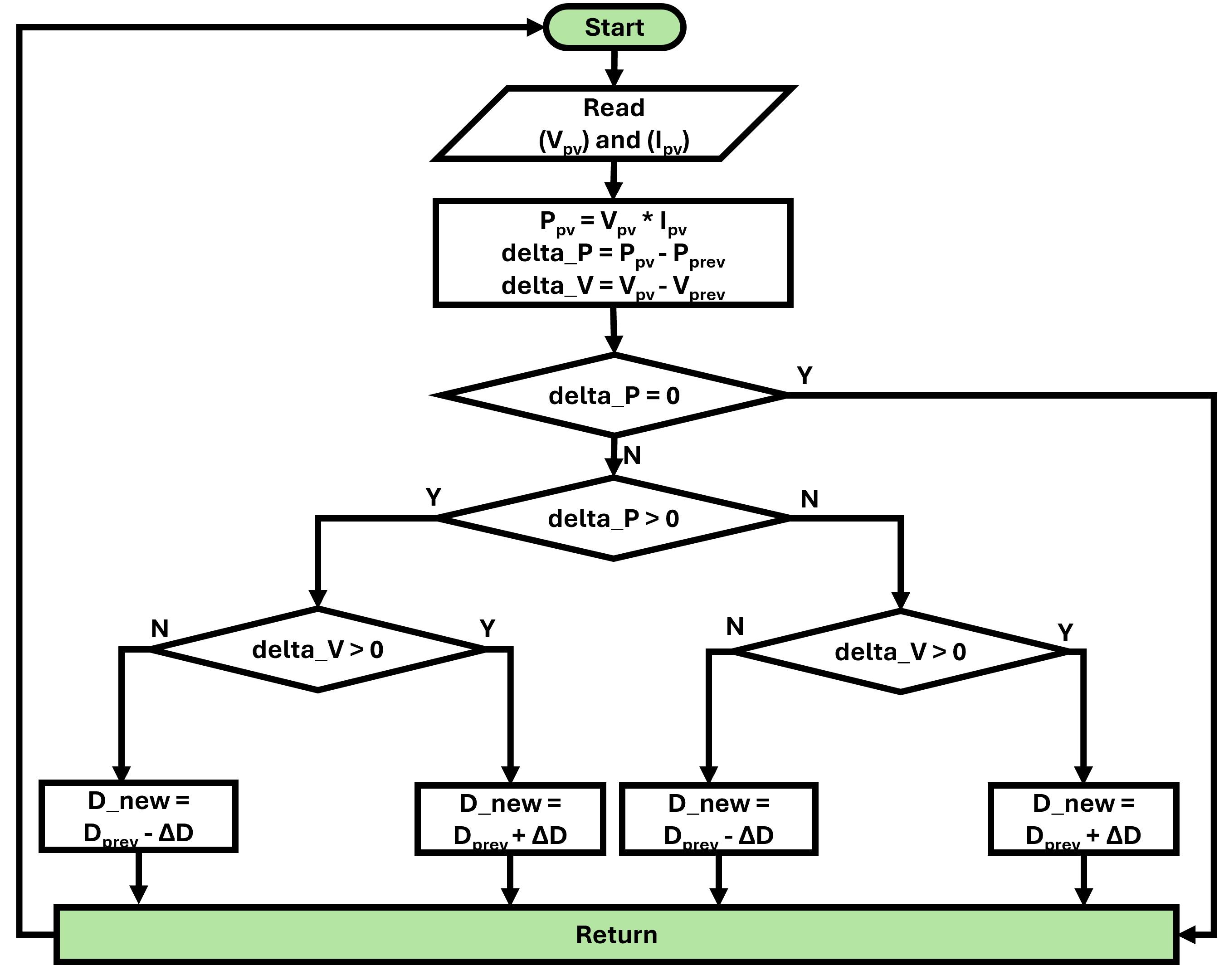}
\caption{Conventional P\&O flowchart.}
\label{fig:conventional P and O flowchart}
\end{figure}

The conventional P\&O algorithm tracks the MPP by incrementally perturbing the duty cycle $D$ and observing the resulting change in power $P=$ $V \cdot I$, where $V$ and $I$ denote the PV voltage and current, respectively. The sign of $\Delta P$ determines the next perturbation direction, as illustrated in Fig.~\ref{fig:conventional P and O flowchart} \cite{ishratguptanayak2024,Alhusseini2024}. Large step perturbations enable fast convergence but cause SS oscillations, whereas small steps reduce oscillations but at the cost of slower response. This is particularly critical for low-power applications, in which any reduction in efficiency drastically degrades device performance.



The persistent trade-off between SS oscillations and tracking speed remains a key limitation of P\&O-based algorithms. To address this, various approaches have been proposed: modified P\&Os \cite{Kahani2023, Mandourarakis2022, RicoCamacho2022}, hybrid methods incorporating fuzzy logic, NN, or metaheuristic optimization \cite{Abdelaziz2015Fuzzy, Zemmit2025, 2024-jjmie, Pham2022FLCPSO, Salman2025, Abouzeid2024}, and logic-based schemes that periodically suspend MPPT operation to suppress oscillations \cite{Jain2019Standby, Bousmaha2022, Yan2021}. However, these strategies often introduce added computational complexity, latency, or the need for offline training, memory, and additional circuitry, limiting their suitability for low-power, real-time systems. These limitations motivate the need for a lightweight, inherently adaptive solution, which will be discussed next.

\subsection{Proposed Adaptive Gradient Descent–Based P\&O MPPT Algorithm with Initialization Routine} \label{subsec:my_AL}



To overcome these limitations, we propose an adaptive MPPT algorithm applying principles of gradient descent (GD). The algorithm adapts the perturbation step size ($\Delta D$) dynamically with the local gradient of the power–voltage curve, $\frac{dP}{dV}$, as given in (\ref{eq:derivative}), according to the proposed flowchart in Fig.~\ref{fig:proposed algorithm}. This adaptation ensures fast convergence and eliminates SS oscillations, key metrics for real-time, efficient tracking in highly dynamic conditions. The approach is lightweight, requires no training, and is suitable for unsupervised implementations.

Gradient descent is a first-order optimization technique that iteratively updates parameters in the direction of the steepest descent in order to maximize or minimize an objective function. When applied in real-time MPPT, it adjusts the duty cycle based upon the $\frac{dP}{dV}$. A positive $\frac{dP}{dV}$ implies operation left of MPP (wherein a decreased duty cycle is needed), and a negative $\frac{dP}{dV}$ implies operation to the right side. This bidirectional, continuous adjustment eliminates the need to
periodically suspend or disable MPPT operation to suppress steady-state oscillations. The update based upon the slope is scaled by a tunable factor $\beta$, which provides a balance between speed and settling accuracy in SS. Its analytic expression is derived below:

\begin{figure}[H]
\centering
\includegraphics[width=3.25in]{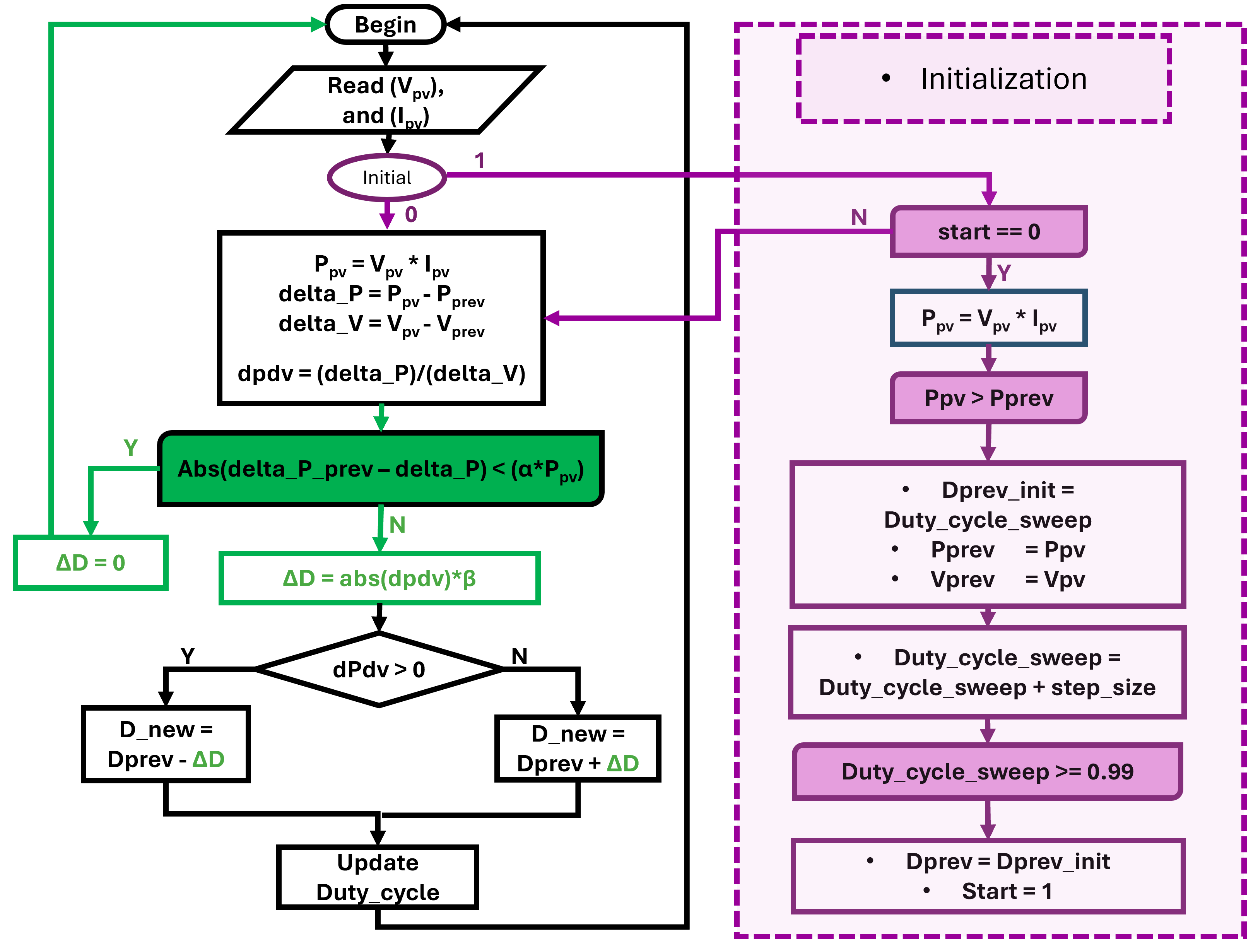}
\caption{Proposed adaptive gradient decent-based P\&O algorithm with initialization routine.}
\label{fig:proposed algorithm}
\end{figure}
\vspace{-8mm}


\begin{equation}
\frac{d P}{d V}=I+V \cdot\left(-\frac{\frac{I_s}{\alpha V_t} e^{\frac{V+I R_s}{\alpha V_t}}+\frac{1}{R_P}}{1+\frac{I_s R_s}{\alpha V_t} e^{\frac{V_t+I R_s}{\alpha V_t}}+\frac{R_s}{R_P}}\right)
\label{eq:derivative}
\end{equation}

\begin{equation}
\begin{aligned}
\frac{\mathrm{d} P}{\mathrm{~d} V}=I_{p h}-I_s\left(\exp \left(\frac{V}{\alpha V_t}\right)-1\right)-
& (\frac{I_s}{\alpha V_t}) \exp \left(\frac{V}{\alpha V_t}\right)
\end{aligned}
\label{eq:slope}
\end{equation}


In order to resolve SS oscillations, the algorithm compares the difference between consecutive power deltas, $|\Delta P(k) - \Delta P(k-1)|$, with a tunable factor ($\alpha$) of Ppv. This adaptive thresholding suppresses unnecessary updates close to the MPP and the risk of oscillation, ensuring robust performance during transients and SS, making it well-suited for highly dynamic and unsupervised low-power PV applications.

A major limitation of GD techniques is their tendency to converge on local extrema in multimodal, non-convex search spaces, affecting real-time performance. This can be mitigated by gradient modifiers like Momentum \cite{Rumelhart1986,MAPO2024} and Adam \cite{kingma2015adam,NEURIPS2024_350e718f} that enhance convergence stability at the cost of extra tunings; advanced initializations like Xavier \cite{glorot2010understanding,Liang2023} and He \cite{he2015delving,Boulila2021} minimize local optima risk but do not guarantee multimodal trap escape; randomness injection like PSO \cite{Plevris_2010,Nature2024,Guo2025} and spawning GD techniques \cite{Sheikhottayefe2025,Shi2025} enhance exploration at the cost of potential instability and slow convergence, metaheuristics, and hybrid techniques like genetic algorithms with GD \cite{Alhijawi2023,SheikhHosseini2022} enhance adaptability in complex landscapes but increase computational load. Despite these advances, applying them in real-time MPPT under PSCs with multiple P-V peaks remains challenging.

To address this issue, we add a lightweight, optional initialization routine, outlined in pink in Fig.~\ref{fig:proposed algorithm}. Through this initialization, the algorithm takes a coarse scan of the duty cycle range and saves the resulting duty cycle that provides the maximum observed power. This single scan has the effect of placing the starting operating point close to the GMPP, lowering the risk of local trapping. The optional initialization routine is designed to be resource-efficient, storing just two power and duty cycle values. It does not impose much additional computational load while improving algorithm robustness under PSC.


\section{SIMULATION RESULTS} \label{section:Simulation}

\begin{figure}[H]
\vspace{-8pt}
\centering
\includegraphics[width=3in]{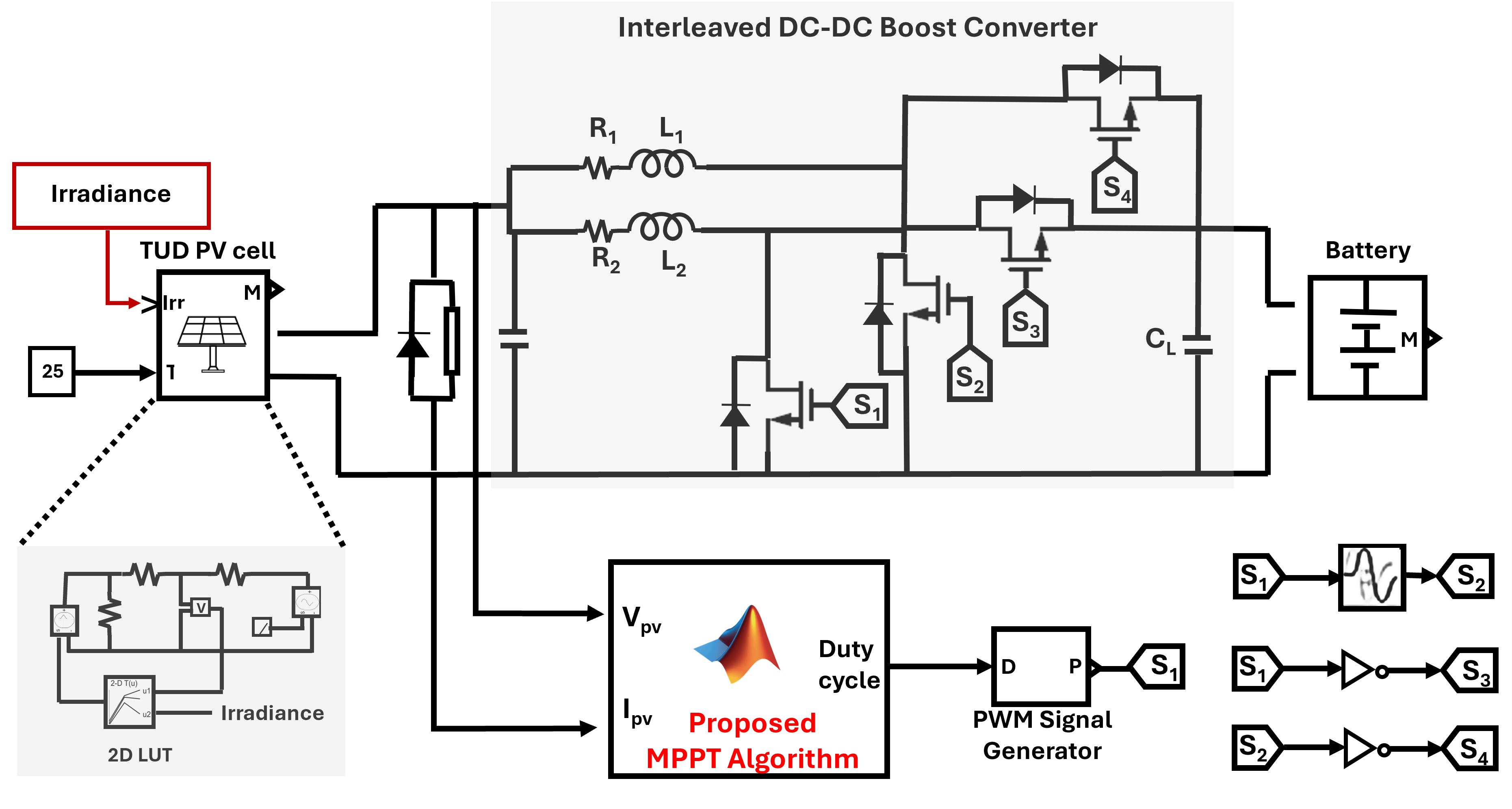}
\caption{Simulink model of the proposed PV system.}
\label{fig:proposed system}
\end{figure}

The proposed MPPT algorithm was verified in MATLAB/Simulink with a parameterized PV model, a 2-phase interleaving boost converter, and a lithium-ion rechargeable cell, as depicted in Fig.~\ref{fig:proposed system}. The load simulates a NeuroLogger 3 probe (25 mA, 3.7 V) powered from a 40 mAh battery for 75 minutes' continuous operation \cite{Ide2022}. PV modeling incorporates high-fidelity characterization from the Photovoltaic Materials and Devices group at Delft University of Technology \cite{Cao2022} under standard AM1.5G irradiance by the parameters given in \autoref{tab:pv cell spec}. To satisfy power and form-factor requirements, the PV area is scaled to 10.49 cm\textsuperscript{2} and gives 250 mW at MPP. The converter’s voltage gain in (\ref{eq:gain}) provides a charging voltage for a 4.2 V lithium-ion cell, while the component specifications given in \autoref{tab:Interleaved components} are for testing the proposed algorithm for operation in a stable manner.

\begin{equation} \label{eq:gain}
G = \frac{V_{\text{out}}}{V_{\text{in}}} = \frac{4.2~\mathrm{V}}{0.650~\mathrm{V}} \approx 6.46
\end{equation}

The average MPPT efficiency in (\%), as a primary evaluation metric in this study is defined in (\ref{eq:efficiency}) \cite{Kahani2023}, where $P_{\mathrm{PV}}(t)$ is the instantaneous harvested power and $P_{\max}(t)$ is the theoretical maximum power, this measure captures transient as well as SS tracking.

\begin{equation}
\eta_{\mathrm{MPPT,avg}} = \frac{\int P_{\mathrm{PV}}(t) \mathrm{d}t}{\int P_{\max}(t) \mathrm{d}t}
\label{eq:efficiency}
\end{equation}

\begin{table}[H]
\centering
\caption{PV Cell Electrical Specifications \cite{Cao2022}.}
\label{tab:pv cell spec}
\footnotesize
\renewcommand{\arraystretch}{0.9}
\begin{tabularx}{\columnwidth}{|l|X|}
\hline
\textbf{Electrical Data at STC} & \textbf{Value} \\
\hline
Peak power $P_{\max}$ & $(93.3 \pm 1.0)$ mW \\
Form factor & $(82.18 \pm 0.58)\%$ \\
Voltage (maximum power) & $(650.35 \pm 1.35)$ mV \\
Current (maximum power) & $(143.5 \pm 1.2)$ mA \\
Open-circuit voltage & $(721.4 \pm 1.8)$ mV \\
Short-circuit current & $(157.4 \pm 1.5)$ mA \\
Cell efficiency & $(23.83 \pm 0.29)\%$ \\
Designated area & $(3.915 \pm 0.020)$ cm$^{2}$ \\
\hline
\end{tabularx}
\renewcommand{\arraystretch}{1.2} 
\end{table}
\vspace{-8pt}

\begin{table}[H]
\centering
\caption{Interleaved Boost Converter Specifications.}
\label{tab:Interleaved components}
\footnotesize
\renewcommand{\arraystretch}{0.9}
\setlength{\extrarowheight}{2pt} 

\begin{tabularx}{\columnwidth}{|l|X|}
\hline
\textbf{Component} & \textbf{Value} \\
\hline
inductance -- ESR & $2.2\,\mu\mathrm{H}$ – 90 m$\Omega$ \\
Capacitor & $10\,\mu\mathrm{F}$ \\
Voltage gain & (4.2/0.65) V \\
Switching frequency & 1 MHz \\
Converter efficiency & 96.43\% \\
Sample time & $1 \times 10^{-8}$ s \\
\hline
\end{tabularx}
\renewcommand{\arraystretch}{1.2} 
\end{table}


The following section evaluates the proposed algorithm across five evaluation metrics: (A) irradiance dynamics tracking, (B) topology-independence and input-source agnosticism, (C) thermal stability, (D) resilience under PSC, and (E) computational complexity evaluation.

\subsection{Dynamic Irradiance Tracking Evaluation}

\begin{figure*}[btp]
\centering
\includegraphics[width=\textwidth]{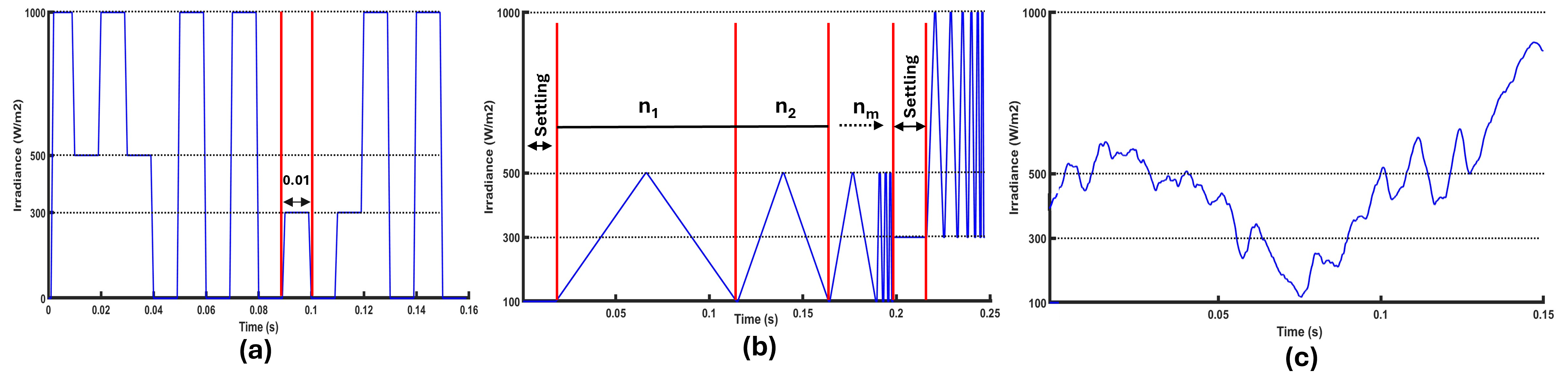}
\caption{Irradiance profiles used for dynamic MPPT performance evaluation: (a) Worst-case step function (Profile 1), (b) EN50530 standard dynamic test (Profile 2), and (c) experimental rodent trajectory-based profile (Profile 3).}
\label{fig:all_irr}
\end{figure*}

Three representative irradiance profiles were developed to evaluate the performance of the proposed adaptive gradient decent-based algorithm, covering abrupt, gradual, and stochastic irradiance changes as depicted in: Fig.~\ref{fig:all_irr}(a) worst-case step function, Fig.~\ref{fig:all_irr}(b) EN50530 standard dynamic test, and Fig.~\ref{fig:all_irr}(c) an experimentally acquired irradiance profile derived from rodent locomotion data.

\begin{figure*}[btp]
\centering
\includegraphics[width=5in]{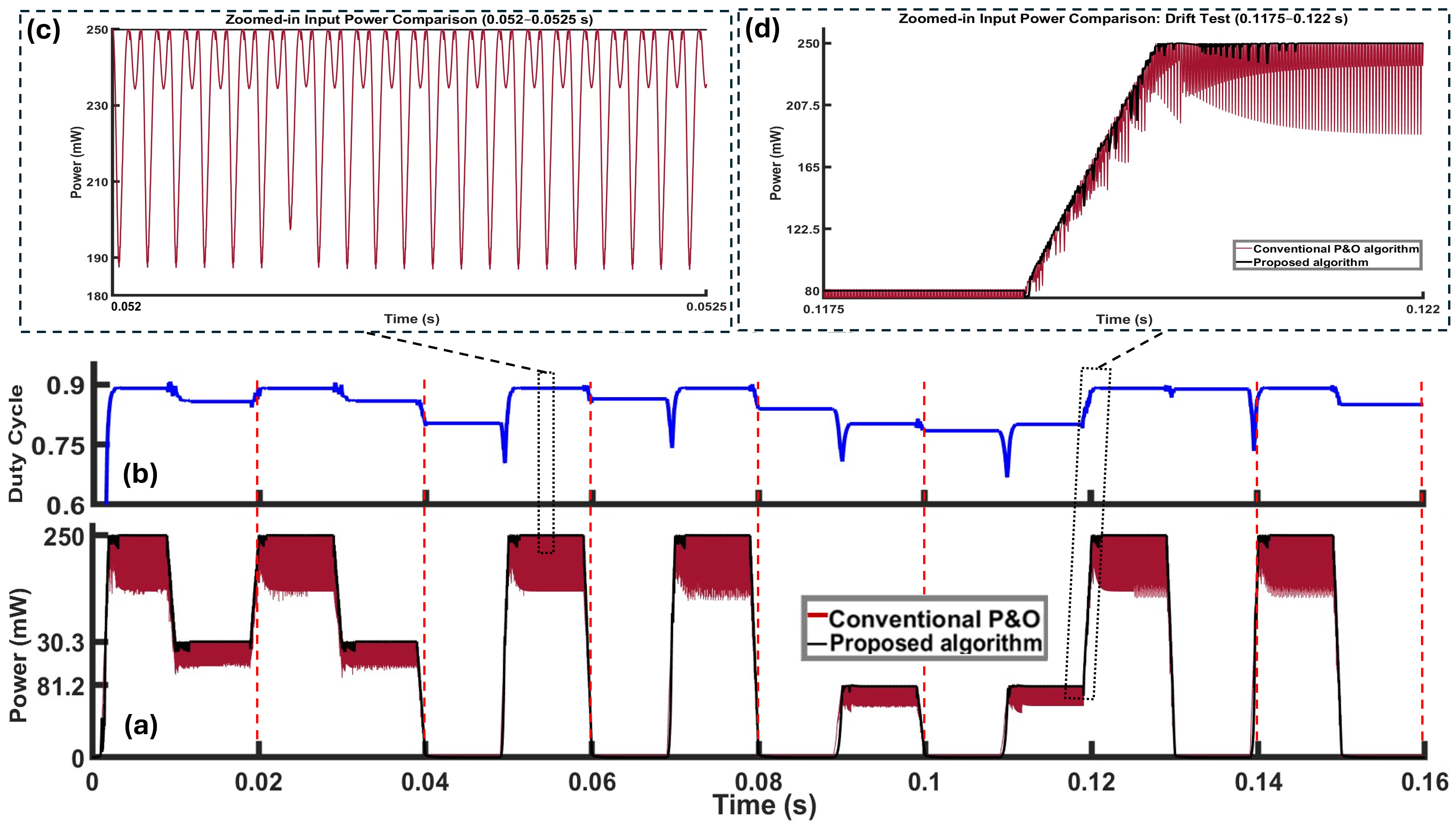}
\caption{Tracking performances of conventional P\&O algorithm and proposed algorithm at Profile 1: (a) Full power response. 
(b) Proposed algorithm's duty cycle. 
(c) Zoomed SS view. 
(d) Zoomed transient view.
}
\label{fig:comparison Irr1}
\end{figure*}

\textbf{Profile 1} is designed to emulate worst-case irradiance changes. It combines abrupt step changes, from complete darkness (0 W/m²) to full irradiance (1000 W/m²) and back to test transient response (drift test). Additionally, SS performance is assessed at multiple irradiance levels to stress the algorithm under both dynamic and static extremes. Fig.~\ref{fig:comparison Irr1} compares the proposed algorithm with the conventional P\&O algorithm under Profile1, evaluated over a 0.16 s at fixed cell temperature $\left(25^{\circ} \mathrm{C}\right)$.


\begin{itemize}
    \item Fig.~\ref{fig:comparison Irr1}(a) shows the resulting tracking performance of the conventional P\&O algorithm in red and the proposed algorithm in black at Profile 1.
    \item Fig.~\ref{fig:comparison Irr1}(b) shows the corresponding duty cycle waveform of the proposed algorithm. By adapting the step size based on the P-V gradient, leading to fast yet stable convergence during both dynamic and SS periods, successfully locked at MPP, suppresses oscillations, and maintains a stable duty cycle.
    \item Fig.~\ref{fig:comparison Irr1}(c) zooms in on the system's SS response, where the proposed algorithm rapidly locks onto the MPP, and meanwhile is still fully awake, while conventional P\&O shows persistent oscillations.

    \item Fig.~\ref{fig:comparison Irr1}(d) zooms in on the transient response, where the proposed algorithm achieves 22× faster tracking with minimal oscillations compared to the modified P\&O in \cite{Kahani2023}, reaching 99.89\% dynamic and 99.94\% STC MPPT efficiency, confirming its robustness under extreme transients.
     
\end{itemize}

\begin{figure*}[btp]
\centering
\includegraphics[width=4.8in]{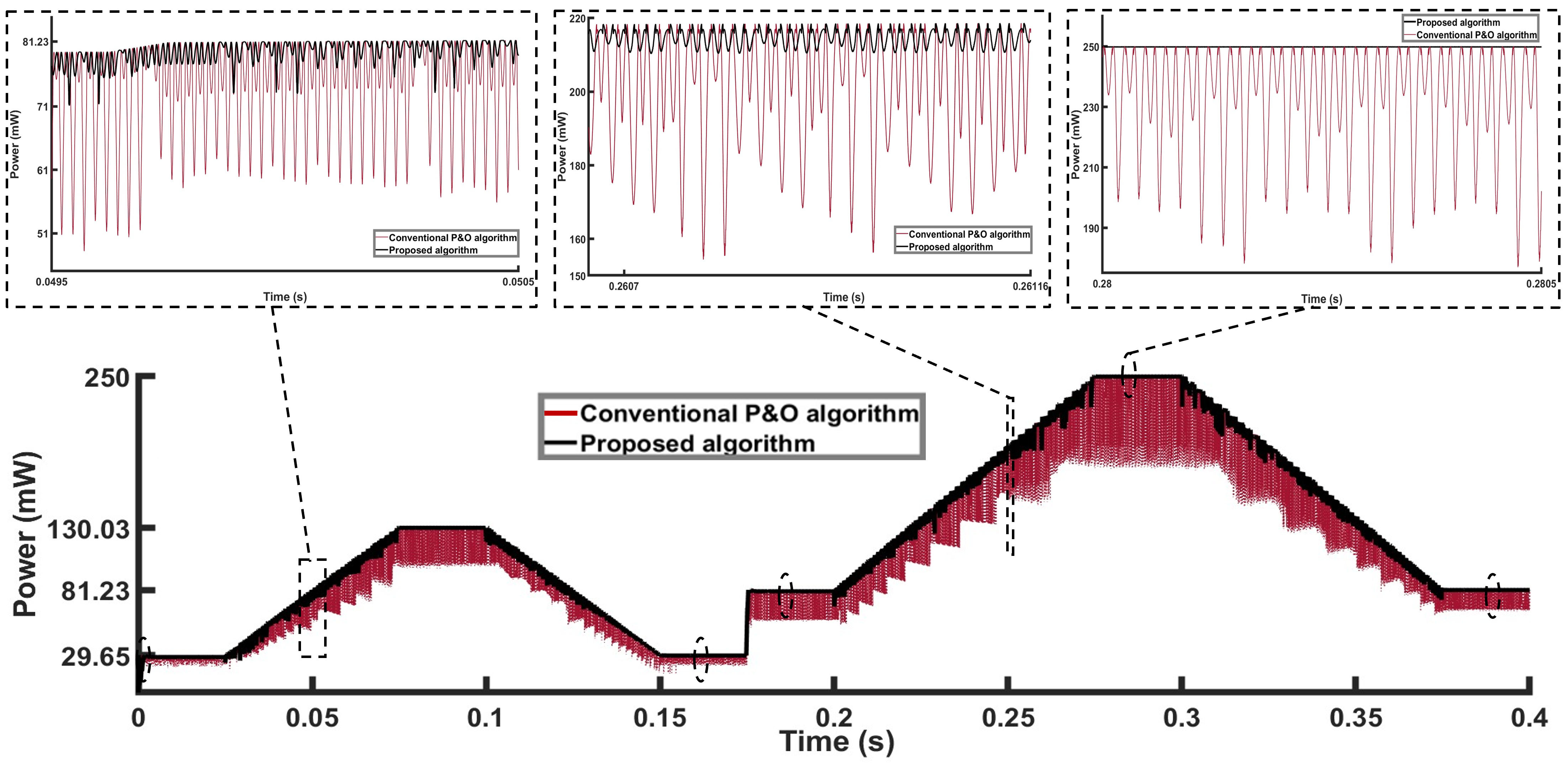}
\caption{Tracking performances of conventional P\&O algorithm and proposed algorithm at Profile 2.}
\label{fig:comparison Irr2}
\end{figure*}



\textbf{Profile 2} conforms to the EN50530 standard dynamic test \cite{EN50530,RicoCamacho2022}, featuring slower ramps for simulating realistic gradual irradiance changes across 0.4 s as illustrated in Fig.~\ref{fig:comparison Irr2}. The proposed algorithm surpasses the conventional P\&O by achieving 8.39\% higher MPPT efficiency at Profile (2). The detailed timing for Profile (2) is outlined in \autoref{tab:EN50530_slow_dynamic_test}.

\begin{table}[H]
\centering
\caption{Profile 2 Test Timings and Results.}
\label{tab:EN50530_slow_dynamic_test}
\resizebox{\columnwidth}{!}{%
\begin{tabular}{|c|c|c|c|c|c|c|c|}
\hline
\textbf{Test Type} & \textbf{Settle} & \textbf{Rise} & \textbf{Rest} & \textbf{Fall} & \textbf{Rest} & \textbf{Conv. P\&O} & \textbf{Proposed} \\
\hline
10\% -- 50\%  & 0.025 & 0.050 & 0.025 & 0.050 & 0.025 & \multirow{2}{*}{90.98} & \multirow{2}{*}{99.37} \\
30\% -- 100\% & 0.025 & 0.075 & 0.025 & 0.075 & 0.025 &                        &                        \\
\hline
\multicolumn{8}{|c|}{\textbf{Total test time: 0.4 s}} \\
\hline
\end{tabular}
}
\end{table}


\textbf{Profile 3} simulates dynamic irradiance conditions recorded during an in vivo experiment with freely moving rodents. With post-processing of a head-mounted 3-axis Intan RHD accelerometer, time-varying light exposure was projected onto a 0.17 s irradiance trace (Fig.~\ref{fig:all_irr}-(C)), simulating stochastic irradiance variations. Profile 3 poses challenges to conventional and intelligent MPPT algorithms owing either to low responsiveness, reliance upon offline training, or pre-known data. The proposed algorithm performance is compared to eight widely adopted MPPT algorithms such as IC, hill climbing (HC), conventional and modified P\&O in  \cite{Kahani2023}, ANN (using $P_{\mathrm{PV}}$ and $P_{\mathrm{PV},\Delta P}$), PSO-tuned FLC \cite{Priyadarshi2023}, and adaptive neuro-fuzzy inference system (ANFIS) \cite{Machesa2022} at Profile 3. As reported in \autoref{tab:mppt_rat_movement}, by achieving 99.21\% MPPT efficiency, the proposed algorithm outperforms all, except ANFIS and PSO-tuned fuzzy logic controllers.

\begin{table}[htbp]
\centering
\caption{MPPT Efficiencies for Irradiance Profile 3.}
\label{tab:mppt_rat_movement}
\begin{tabular}{|l|c|}
\hline
\textbf{MPPT Algorithm} & \textbf{Efficiency (\%)} \\
\hline
Conventional P\&O & 89.46 \\
Incremental conductance (IC) & 94.45 \\
Hill climbing (HC) & 92.44 \\
Kahani's modified P\&O & 50.76 \\
Neural Network (P) & 99.25 \\
Neural Network (P, $\Delta \mathrm{P}$) & 98.43 \\
Fuzzy tuned with PSO & 99.51 \\
Adaptive neurofuzzy inference system (ANFIS) & 99.57 \\
\textbf{Proposed algorithm} & \textbf{99.21} \\
\hline
\end{tabular}
\vspace{-10pt}
\end{table}

\begin{figure}[!ht]

\centering
\includegraphics[width=3.22in]{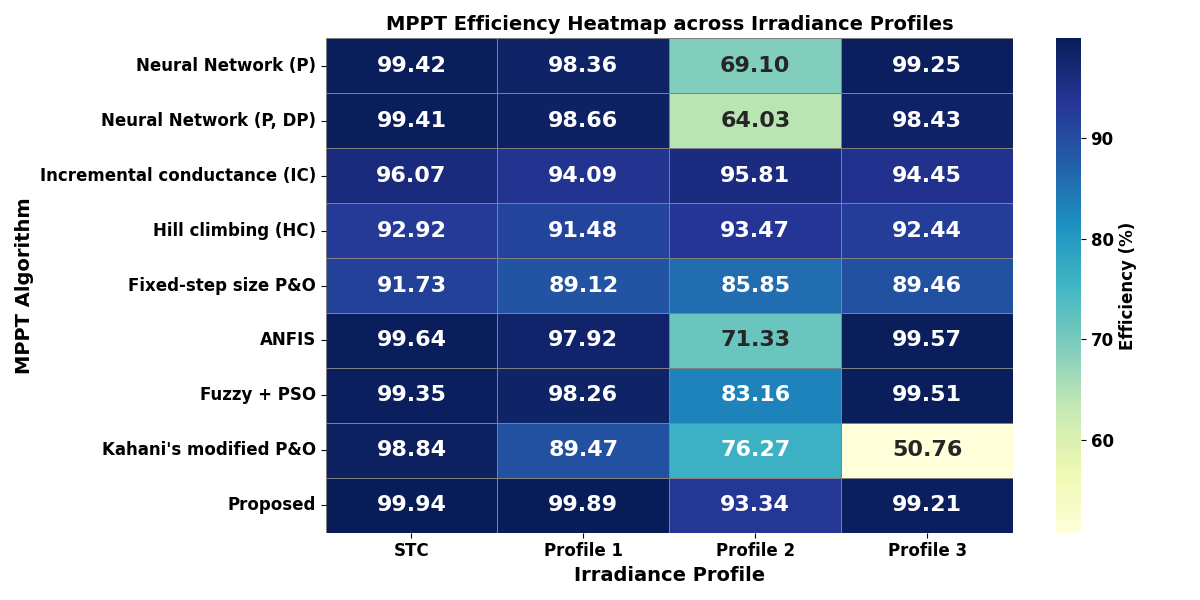}
\caption{MPPT efficiency heatmap comparing the proposed algorithm with conventional, intelligent, and hybrid algorithms across: STC, Profiles 1, 2, and 3, respectively.}
\label{fig:Heatmap}
\vspace{-10pt}
\end{figure}

 A consolidated heatmap (Fig.~\ref{fig:Heatmap}) compares all eight mentioned algorithms for the described 3 irradiance profiles. The proposed algorithm achieves the highest MPPT efficiencies while remaining low in computational load, as opposed to intelligent or hybrid ones. Such performance-complexity trade-offs, explored further in \autoref{subsec:computational_load}, prove the algorithm's suitability for real-time and low-power embedded PV systems.\vspace{-10pt}

\subsection{Evaluation of Converter and PV Model Independence}


\arrayrulecolor{black}
\setlength{\arrayrulewidth}{0.3mm}

\begin{table*}[htbp]
\caption{Converter Topologies Evaluated with the Proposed Algorithm: Characteristics and Efficiency. Color-code: green shades represent non-isolated (light to dark = increasing complexity), blue shades indicate isolated topologies.}
\label{tab:merged_converter_table}
\centering
\renewcommand{\arraystretch}{2}
\resizebox{\textwidth}{!}{
\begin{tabular}{|c|c|>{\centering\arraybackslash}p{2.5cm}|>{\centering\arraybackslash}p{2.5cm}|>{\centering\arraybackslash}p{2.5cm}|>{\centering\arraybackslash}p{2.5cm}|}
\hline
\textbf{DC-DC Converter} &
\textbf{Ideal voltage Gain in CCM} &
\textbf{Converter Efficiency (\%)} &
\textbf{MPPT Efficiency at STC (\%)} &
\textbf{MPPT Efficiency at Irradiance Profile 1 (\%)} &
\textbf{MPPT Efficiency Using Output Specs (\%)} \\
\hline
\rowcolor{noniso1}
Simple Boost & $\frac{1}{1-D}$ & 93.16 & 99.76 & 97.11 &99.56 \\
\hline
\rowcolor{noniso2} Buck-Boost & $\frac{-D}{1-D}$ & 91.02 - 91.32 & 99.53 & 84.78 & 99.78 \\
\hline
\rowcolor{noniso3} Bidirectional Buck-Boost & $\left\{\begin{array}{l}\frac{1}{1-D} \, \text{(Boost)} \\ \frac{1}{D} \, \text{(Buck)}\end{array}\right.$ & 94.11 & 98.84 & 96.19 & 92.84 \\
\hline
\rowcolor{noniso4} Cuk & $\frac{-D}{1-D}$ & 89.44 & 98.31 & 85.86 & 99.89 \\
\hline
\rowcolor{noniso5} SEPIC & $\frac{D}{1-D}-\frac{(V_f + V_{switch})}{V_i}$ & 92.52 & 96.89 & 89.20 & 99.81 \\
\hline
\rowcolor{noniso6} Zeta & $\frac{D}{1-D}$ & 91.68 & 98.26 & 89.71 & 99.86 \\
\hline
\rowcolor{noniso7} Quadratic Boost & $\frac{1}{(1-D)^2}$ & 89.64 & 98.86 & 88.07 & 97.95 \\
\hline
\rowcolor{noniso8} Multilevel Boost & $\begin{aligned} & V_{o, N}=N . V_{i n}-\sum V_{f, N} \\ & N \cdot\left(\frac{1}{1-D}\right)\end{aligned}$ & 92.11 & 99.16 & 95.01 & 98.96 \\
\hline
\rowcolor{noniso9} High Step-Up Boost & $1 + \frac{n}{1-D}$ & 93.75 & 98.64 & 92.07 & 99.84 \\
\hline
\rowcolor{noniso10} Switched Capacitor (Fibonacci) & $\begin{aligned} & F_{k+1} ; \\ & F_j=\left(\varphi^j-(1-\varphi)^j\right) / \sqrt{5}, \\ & \varphi=(1+\sqrt{5}) / 2\end{aligned}$ & 95.75 & 98.88 & 88.24 & 98.93 \\
\hline
\rowcolor{noniso11} Interleaved Switched-Capacitor (Hybrid) & $\frac{1}{1-D}+\alpha; \alpha: SC-gain $ & 94.08 & 97.87 & 90.31 & 98.03 \\
\hline
\rowcolor{noniso12} 2-Phase Interleaved Boost & $\frac{1}{1-D}$ & 96.43 & 99.79 & 99.89 & 97.23 \\
\hline
\rowcolor{iso1} Flyback & $\frac{nD}{1-D}$ & 92.76 & 97.64 & 75.25 & 97.61 \\
\hline
\rowcolor{iso2} Forward & $nD$ & 90.23 & 94.52 & 83.38 & 94.51 \\
\hline
\rowcolor{iso3} Push-Pull & $2nD$ & 93.64 & 96.41 & 82.71 & 94.11 \\
\hline
\rowcolor{iso4} Half-Bridge & $nD$ & 95.80 & 99.19 & 86.78 & 99.36 \\
\hline
\rowcolor{iso5} Full-Bridge & $2nD$ & 86.56 & 99.57 & 92.01 & 99.43 \\
\hline
\rowcolor{iso6} Resonant &
$\left(\frac{N_s}{N_p}\right)\cdot M_{\text{Resonant-tank gain}}$ &
97.13 & 97.52 & 73.22 & 93.69 \\
\hline
\end{tabular}}

\begin{minipage}{\linewidth}
\footnotesize
\vspace{1mm}
\textit{Note:} $D$: duty cycle, $V_{\text{in}}$: input voltage, $V_f$: diode forward voltage, $n$: turns ratio, $N_p$, $N_s$: primary/secondary windings, $M$: resonant gain factor.
\end{minipage}
\end{table*}



The hardware independence of the proposed algorithm was validated across a wide range of DC-DC converter topologies operating in continuous conduction mode (CCM), including non-isolated and isolated configurations, covering a broad spectrum of voltage gains, efficiencies, and implementation complexities. These topologies are summarized in \autoref{tab:merged_converter_table}, where converter types are color-coded by complexity. Ideal voltage gain formulas were provided with respect to the conventional models as detailed in \cite{Forouzesh2017}.

In spite of wide variations in converter topology and dynamic behavior, the proposed algorithm consistently achieves MPPT efficiencies above 98\% at STC and exceeds 95\% at Profile 1. Its highest performance of 99.89\% is observed when paired with a 2-phase interleaved boost converter. As shown in \autoref{tab:merged_converter_table}, ideal voltage gains and efficiency metrics demonstrate the algorithm’s robustness across diverse power stage configurations. Notably, no topology-specific tuning or parameter adaptation is required, highlighting the algorithm’s converter independence. To further validate this, the algorithm was tested using load-side voltage and current measurements instead of direct PV-side sensing, an important practical constraint in integrated PV systems. This substitution had a negligible impact on performance, further confirming the algorithm’s resilience to both electrical architecture and sensing configuration. These results collectively support the proposed algorithm’s broad applicability in dynamic, topology-agnostic energy harvesting systems.


\subsection{MPPT Performance Under Temperature Variations}



\begin{figure}[H]
\centering
\includegraphics[width=2.5in]{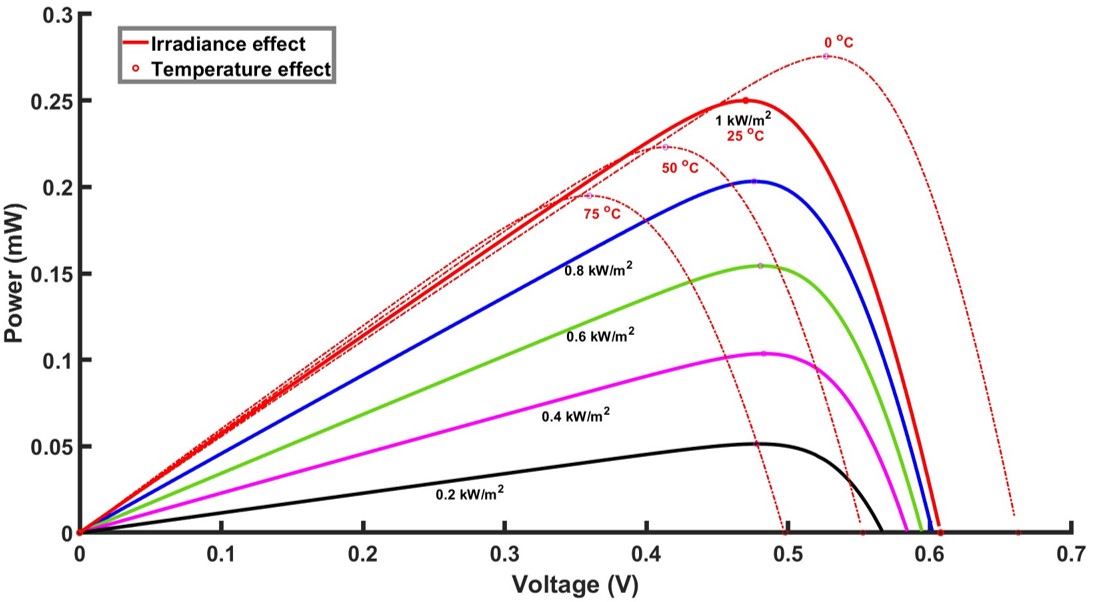}
\caption{Temperature variation effect on the P-V curve.}
\label{fig:temp effect}
\vspace{-10pt}
\end{figure}

While most MPPT research focuses on irradiance dynamics and PSC, temperature variations also significantly affect PV performance; their impact on the P–V curve is depicted in Fig.~\ref{fig:temp effect} \cite{Chellakhi2021}.

Thermal stability of the proposed algorithm is tested under static and dynamic temperature tests according to the EN50530, IEC 61853-1, and Sandia standards \cite{IEC61853_1,IEC62509}. \autoref{tab:temp_variation_results} presents the proposed algorithm’s efficiencies and tracking times, with all cases exceeding 99.8\% MPPT efficiency except at 75 degrees. Tracking time refers to the duration from a step change to the point where output power stabilizes within ±2\% of the new MPP \cite{Teklehaimanot2024,HussaianBasha2020}. These results demonstrate thermal stability of the proposed algorithm.\vspace{-10pt}

\begin{table}[H]
\caption{Temperature Tests Description and Results.}
\label{tab:temp_variation_results}
\centering
\renewcommand{\arraystretch}{1.2}
\resizebox{\columnwidth}{!}{
\begin{tabular}{|p{3.5cm}|p{3.5cm}|p{3.5cm}|}
\hline
\centering\arraybackslash \textbf{Temperature Variation Scenarios} & 
\centering\arraybackslash \textbf{MPPT Efficiency (\%)} & 
\centering\arraybackslash \textbf{Tracking Time (s)} \\
\hline
\multicolumn{3}{|c|}{\textbf{Static Temperature Tests}} \\
\hline
0°C & 99.8632 & 196.234 $\mu$s \\
25°C (STC) & 99.8605 & 225.162 $\mu$s \\
50°C & 99.8418 & 226.110 $\mu$s \\
75°C & 99.6763 & 307.369 $\mu$s \\
\hline
\multicolumn{3}{|c|}{\textbf{Dynamic Temperature Tests}} \\
\hline
20°C $\rightarrow$ 50°C (ramp) & 99.9598 & - \\
25°C $\rightarrow$ 45°C (ramp) & 99.9349 & - \\
\hline
\end{tabular}
}
\end{table}

\subsection{MPPT Performance Evaluation Under PSC}

\begin{figure}[!ht]
\vspace{-8pt}
\centering
\includegraphics[width=3.2in]{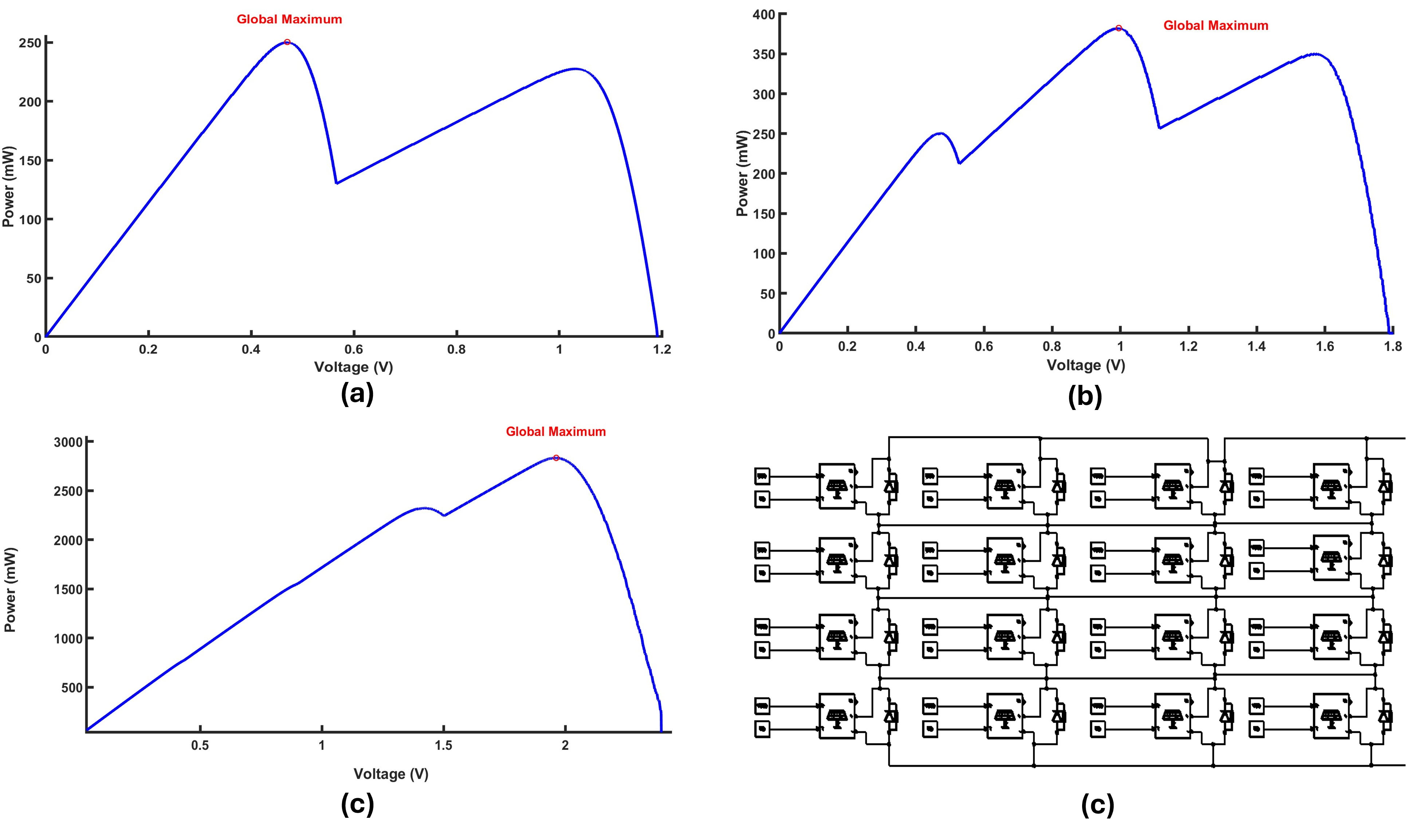}
\caption{Defined PSC scenarios: (a) Two-Peak, (b) Three-Peak, and (c, d) TCT configurations.}
\label{fig:PSC}
\end{figure}

Partial shading conditions create multiple local maxima in the P–V curve, which challenge conventional, intelligent, and gradient-based MPPT algorithms. Five challenging PSC scenarios were used to evaluate the proposed algorithm's global tracking performance, including \textbf{Moderate} and \textbf{Strong} scenarios from \cite{Kahani2023} consisting of three parallel
strings with five modules connected in series per string irradiated by 700/300/100 and 750/150/100 W/m² levels, respectively, and 3 developed scenarios as depicted in Fig.~\ref{fig:PSC} including \textbf{Two Peaks} with a GMPP leftside of a secondary peak; \textbf{Three Peaks} with the GMPP between two maxima, and \textbf{total cross tied (TCT)}, which comprised five four-module strings connected in a mesh, subject to irradiance levels 500, 300, 100, 200, and 25 W/m\textsuperscript{2}, producing the most complex shading and resulting in a highly non-convex P–V curve. The proposed approach is evaluated with and without an optional initialization routine. As demonstrated in \autoref{tab:PSC_efficiency}, initialization enhanced MPPT efficiency for all PSC scenarios, boosting MPPT efficiencies by 7.8\% from 91.92\% (Three-Peaks PSC) up to 99.68\%. The algorithm sustained robust tracking despite severe mismatch and configuration complexity, while STC performance remained unaffected, confirming its robustness in PSC conditions.

\begin{table}[H]
\caption{MPPT Efficiency Comparison Under PSC without and with Initialization Routine.}
\centering
\renewcommand{\arraystretch}{1.2}
\resizebox{\columnwidth}{!}{
\begin{tabular}{|l|c|c|}
\hline
\textbf{PSC Scenario} &
\begin{tabular}[c]{@{}c@{}}\textbf{MPPT Efficiency (\%)}\\ \textbf{Without Initialization}\end{tabular} &
\begin{tabular}[c]{@{}c@{}}\textbf{MPPT Efficiency (\%)}\\ \textbf{With Initialization}\end{tabular} \\
\hline
Two-Peaks PSC & 93.66 & 99.39 \\
Three-Peaks PSC & 91.92 & 99.68 \\
 Moderate PSC \cite{Kahani2023} & 94.82 & 99.30 \\
 Strong PSC \cite{Kahani2023}& 96.03 & 99.63 \\
TCT Connection PSC & 97.63 & 99.51 \\
STC & 99.30 & 99.66 \\
\hline
\end{tabular}
}
\label{tab:PSC_efficiency}
\end{table}

\subsection{Computational Load Evaluation} \label{subsec:computational_load}



To be able to compare the complexity of a hardware implementation of the proposed algorithm, a computational complexity analysis is conducted. This evaluation includes all MPPT algorithms studied in this work and some state-of-the-art P\&O-based algorithms, focusing on low-power VHDL-compatible digital IC implementation. A normalized ASIC-based gate-level cost measure $X$ is defined, such that 1$X$ is the computational cost of a 10-bit add operation. Other operations scaled relatively to $X$ are given in synthesis-informed ASIC design heuristics from literature, as given in \autoref{tab:comp_load} \cite{rabaey2003digital,weste2010cmos,smith1997asic,Zeevi2020}.



\begin{figure}[!ht]
\vspace{-8pt}
\centering
\includegraphics[width=3.3in]{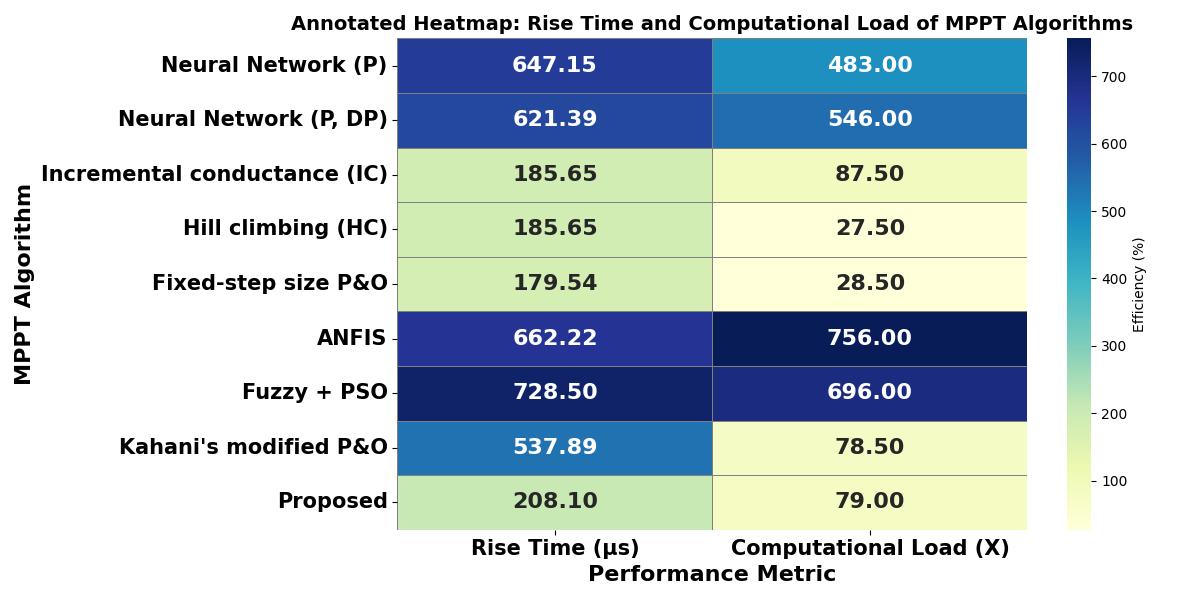}
\caption{Heatmap comparing rise time and normalized computational load of MPPT algorithms.}
\label{fig:Heatmap_comp}
\end{figure}

 Complexity and tracking time for the eight discussed MPPT techniques are compared in Fig.~\ref{fig:Heatmap_comp}. Each computational load is computed per algorithm's execution cycle from its proposed flowchart and mapped to $X$. Intelligent algorithms have the highest complexity due to matrix operations, fuzzy inference, and nonlinear transforms: ANN has 10 hidden neurons that use $\texttt{tansig}$ activation and normalization; ANFIS uses seven $\texttt{psigmf}$ membership functions that use weighted defuzzification; PSO-Fuzzy has optimization overhead. While accurate, these impose substantial hardware costs, restricting use in resource-constrained embedded systems. The proposed algorithm has a low computational cost through gradient estimation and adaptive step sizing.

\vspace{-8pt}

\section{Unified Performance Metric for MPPT Algorithms}

\begin{table*}[htbp]
\caption{Normalized Computational Cost for Two 10-bit Digital Operations (ASIC-Based) \cite{rabaey2003digital,weste2010cmos,smith1997asic,Zeevi2020}.}
\centering
\resizebox{\textwidth}{!}{
\begin{tabular}{|l|l|c|p{6.8cm}|}
\hline
\textbf{Operation} & \textbf{Description} & \textbf{Relative Complexity (X)} & \textbf{Notes} \\
\hline
$A + B$ & 10-bit addition & 1$\times$ & Single-cycle adder; linear with bit-width \\
$A - B$ & 10-bit subtraction & 1$\times$ & Same as addition (add + invert + carry-in) \\
$A \times B$ & 10-bit multiplication & 8--10$\times$ & Array/tree multiplier; grows quadratically \\
$A \div B$ & 10-bit division & 30--40$\times$ & Iterative, slow; typically avoided in hardware \\
$A \gg n$ & Bitwise shift right by $n$ bits & $\sim$0.2$\times$ & Wiring only; negligible logic delay \\
$A \,\&\, B$, $A\,|\,B$ & Bitwise AND, OR & $\sim$0.1$\times$ & Implemented as simple gates \\
$A == B$ & Equality check & $\sim$0.5$\times$ & Uses XOR followed by NOR \\
$A > B$ & Comparator & $\sim$1.5$\times$ & Uses subtractor logic \\
\texttt{if (A > B)} & Control flow (comparator + MUX) & $\sim$2$\times$ & Complexity depends on control logic and branching \\
LUT access & Lookup table (7-bit address) & $\sim$2$\times$ & $\sim$128-entry ROM; 1-cycle latency assumed \\
RAM read/write & 1-word access & $\sim$3$\times$ & Depends on RAM interface and access width \\
$\exp(x)$ & Exponential & 20--40$\times$ & Typically implemented via LUT or CORDIC \\
$\log(x)$ & Logarithm & 20--40$\times$ & Typically LUT or iterative approximation \\
\hline
\end{tabular}
\label{tab:comp_load}
}
\end{table*}

\begin{table*}[htbp]
\caption{Comparison of P\&O-Based MPPT Algorithms in Terms of Efficiency, Dynamics, Complexity, and FoM.}
\centering
\resizebox{\textwidth}{!}{
\begin{tabular}{|c|c|c|c|c|c|c|c|}
\hline
\textbf{Ref.- Year} & \textbf{MPPT algorithm} & \textbf{MPPT efficiency (\%)} & \textbf{Power oscillation (\%)} & \textbf{Power level (W)} & \textbf{Tracking time (s)} & \textbf{Computational load (X)} & \textbf{FoM (\%/s.X)} \\
\hline
\cite{Chouiekh2025} - 2025     & H-ABC-P\&O & 99.85 & 0.15 & 2000                   & 0.14                       & 10413 & 0.068 \\
\hline
\cite{Nagadurga2025} - 2025   &  ChOA optimized P\&O  & 99.82          & 0.18 & 440.87                    & $< 0.002$                      & 1713 & 29.084
\\
\hline
\cite{Nagadurga2025} - 2025   & TLBO optimized P\&O  & 98.86            & 1.14 & 438.63                   & 0.05                      & 2238 & 0.874
 \\
\hline
\cite{Djilali2025} - 2025      & VSS-P\&O  &  99.98 &  0.02 &      62.21   & 0.213                     & 75.5 & 6.216\\
\hline
\cite{Djilali2025} - 2025 & Modified VSS-MPPT-P\&O & 99.98        & 0.01 & 62.21                 & 0.032   & 317 & 9.855
\\
\hline
\cite{Ashwini2025} - 2025  & PSO optimized P\&O & 99.85           & 0.148 & 235.75                     & 0.042                      & 1591.5 & 1.492
 \\
\hline
\cite{Ashwini2025} - 2025   & Adjustable CSA             & 99.98             & 0.018 & 237.23                   & 0.06                       & 184.5 & 9.030
\\
\hline
\cite{Karimi2025} - 2025 & Hybrid-Fuzzy-Based P\&O  & 99.98             & 0.02 & 34.88                    & 5.02                      & 751.5 & 0.026
 \\
\hline
\cite{Ali2025} - 2025 & Hybrid FLC + ANN & 99.97    & 0.079 & 252.4                   & 0.005                      & 802.5 & 24.895
 \\
\hline
\cite{Ali2025} - 2025 & ANN-based & 99.89    & 0.079 & 252.2                   & 0.015                      & 161 & 41.330
 \\
\hline
\cite{Kavuru2024} - 2024 & drift-free P\&O  & 99.9    & 0.0061 & 1640                   & 2.00                      & 69 & 0.724
\\
\hline
\cite{Kavuru2024} - 2024 & Modifed drift-free IC  & 99.9     & 0.061 & 1640                   & 2.10                      & 89.5 & 0,531
 \\
\hline
\cite{CS2024} - 2024 & P\&O based DISMC  & 93.2    & 0.49 & 412.38                   & 0.32                      & 178.5 & 1.624
 \\
\hline
\cite{Zubair2024} - 2024 & Hybrid MPPT with ANN  & 99.20    &  0.72 & 85380                  & 0.23                      & 177 & 2.419
 \\
\hline
\cite{VietAnh2024} - 2024 & Modified P\&O for PSC  & 99.99   & 0.011 & 89.21                   & 0.136                      & 218 & 3.372
\\
\hline
\cite{Wang2024} - 2024 & Hybrid POA \& PO   & 99.97    &  0.03 & 330.4                   & 0.3                    & 165 & 2.019
  \\
\hline
\cite{Hendy2024} - 2024 & MPSO-MPC   & 99.97   & 0.03 & 180.8                  & 0.35                      & 344 & 0.830
 \\
\hline
\cite{Iyer2024} - 2024 & Binary Search-based adaptive P\&O   & 98.47  & 1.53 & 25.8    & 10                      & 90.5  & 0.107
\\
\hline
\cite{Baso2024} - 2024 &  Hybrid IGWO-P\&O   & 99.37    & 0.63 & 137.69                  & 0.15                      & 6025 & 0.109
 \\
\hline
\cite{Naser__2024} - 2024 &  Modified P\&O   & 99.81   & 0.19 & 48.84                  & 0.40                      & 111.5 & 2.234
\\
\hline
\cite{Alhusseini2024} - 2024 &  Fuzzy-Based Adaptive P\&O   & 99.99   & 0.01 & 60.20                 & 0.1                     & 99 & 10.099
 \\
\hline
\cite{HussaianBasha2024} - 2024 &  Hybrid GWO-ANFIS   & 97.22   & 2.78 & 473.40                & 0.031                     & 6677 & 0.457
 \\
\hline
\cite{IchimBurlacu2023} - 2023 &  Fuzzy logic-based   & 99.5   &  0.05 & 280               & 0.2                   & 125 & 3.978 \\
\hline
\cite{NzoundjaFapi2023} - 2023 &  Fuzzy logic-based enhanced P\&O  & 99.6   & 0.05  & 100               & 0.01                   & 340 & 29.279
 \\
\hline
\cite{Huang2023} - 2023 &  IP\textit{SO-WOA }$+$ P\&O  & 99.24   & 0.76 & 5731               & 0.022                   & 158.5 & 28.245
\\
\hline
    \cite{Jabbar2023} - 2023 &  Modified P\&O  & 99.8   & 0 & 160-180               & 0.3                   & 188.5 & 1.765
 \\
\hline
    \cite{YKSEK2023} - 2023 &  P\&O-based variable stepsize  & 99.99   & 0.26 & 160.1           & 0.197                 & 36.5 & 13.870
\\
\hline
    \cite{GM2021} - 2021 &  CPPO  & 99.88    & 0.24 & 250.3          & 0.12                 & 76 & 10.926
 \\
\hline
    \cite{Etezadinejad2022} - 2022 &  Improved and Fast  & 99.8   & 0.1 & 114.5          & 0.006                 & 380 & 43.728
 \\
\hline
    \cite{Manna2021} - 2021 &  Updated P\&O  & 99  & 0.33 & 211.5         & 0.12                 & 74 & 11.112
\\
\hline
    \cite{Rao2021} - 2021 &  Improved P\&O (IP\&O)  & 95.2   & 0.5 & 200         & 0.02                 & 47.500 & 99.712
\\
\hline
    \cite{Swaminathan2022} - 2022 &  Fixed Zone P\&O (FZPO) & 97.85  &  0.2 & 200         & 4                 & 64.500 & 0.379
\\
\hline
    \cite{Geetha2021} - 2021 &  DSSO P\&O & 99.99   & 0 & 300.04         & 0.041                & 157.5 & 15.484
 \\
\hline
    \cite{Kahani2023} - 2023 & Kahani's Modified P\&O & 98.21 & 0.65 & 2500         & 0.05-0.1                & 78.5 & 12.430
 \\
\hline
     \rowcolor{noniso5}  2025 &  Proposed & 99.89  & 0 & 0.01         & 0.1702  & 79 & 126.443
 \\
\hline

\end{tabular}
\label{tab:mppt_compare}
}
\end{table*}


To enable hardware-relevant comparison between the proposed and state-of-the-art P\&O-based MPPT algorithms, a benchmarking figure of merit (FoM) is proposed in (\ref{eq:FoM}). This FoM focuses on tracking performance but penalizes time, computational cost, and power oscillation. In contrast to analog-focused metrics (e.g., Schreier FoM \cite{Manganaro2011,Manganaro2018}), there is no inherent proportionality between these merits in nonlinear PV systems. Hence, the proposed FoM is designed to emphasize empirical trade-offs and interpretability, rather than analytical rigidity. The system's power level is excluded for maintaining application-independent fairness across use cases.

\begin{equation}
\mathrm{FoM}=\frac{\eta_{\mathrm{MPPT}}}{\left(T_{\text {track }} \cdot X_{\text {comp }}\right) \cdot\left(1+\frac{\Delta P_{\mathrm{ss}}}{100}\right)}
\label{eq:FoM}
\end{equation}

\begin{figure*}[!ht]
\vspace{-8pt}
\centering
\includegraphics[width=\textwidth]{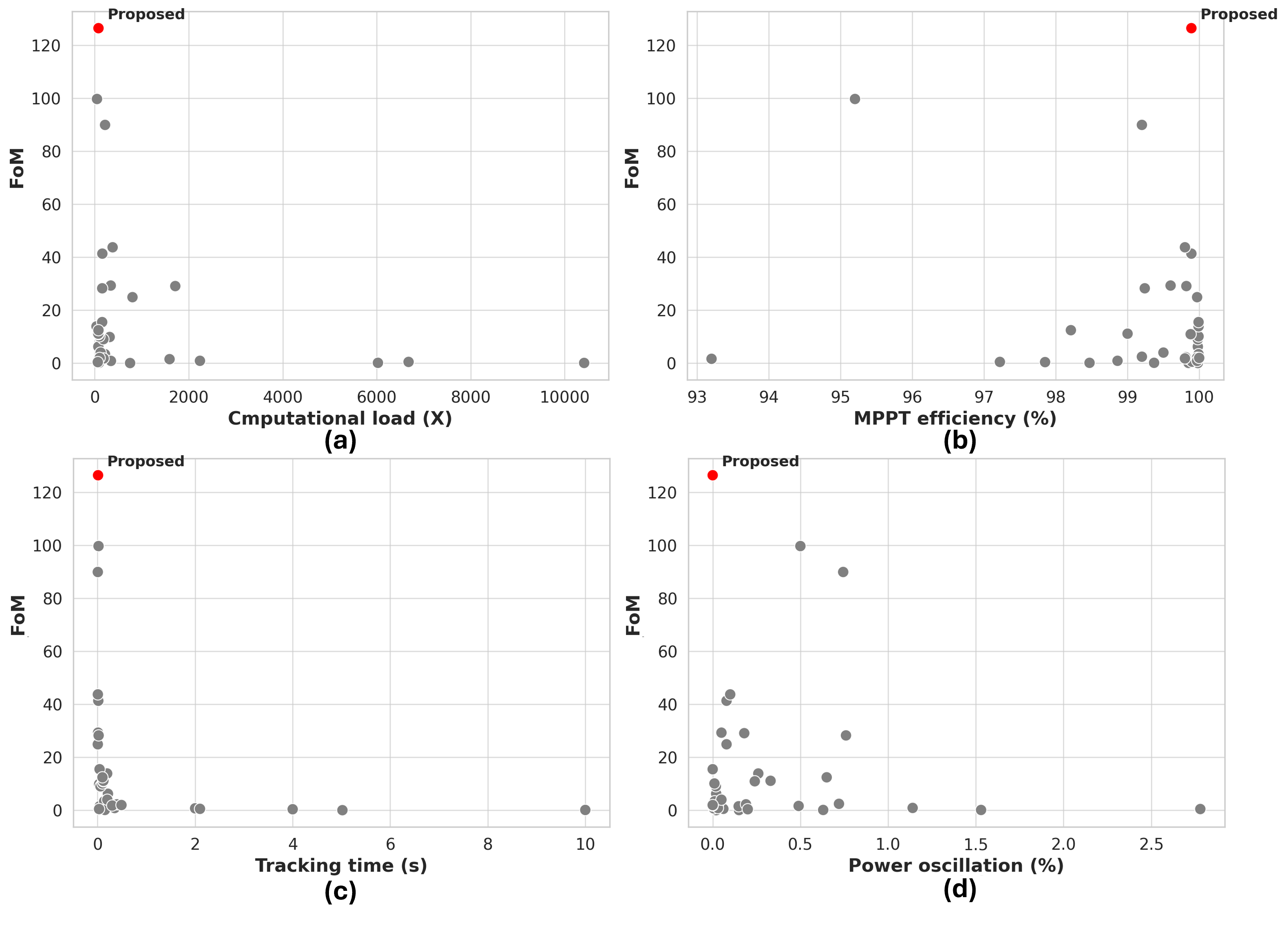}
\caption{Scatter plot of the proposed FoM versus (a) computational complexity, (b) MPPT efficiency, (c) Tracking time, (d) Power oscillation across reviewed algorithms in \autoref{tab:mppt_compare}; proposed method highlighted in red.}
\label{fig:FoM_Comparison}
\end{figure*}

 \autoref{tab:mppt_compare} presents a comparative summary of 35 state-of-the-art P\&O-based MPPT algorithms, including MPPT efficiency, steady-state power oscillation, power level, tracking time, and normalized computational load. The final column introduces the FoM, derived from these parameters based on \autoref{eq:FoM}, to enable fair comparison based on MPPT efficiency, tracking performance, and implementation cost. The corresponding scatter plots in Fig.~\ref{fig:FoM_Comparison} visualize the resulting FoM versus each key metric; the good performance of the proposed algorithm can be appreciated from various performance angles.

\titlespacing{\section}{0pt}{*1}{*1}
\section {DISCUSSION AND CONCLUSION}


This study introduces an adaptive gradient descent–based P\&O MPPT algorithm tailored for highly dynamic low-power OWPT systems. By adapting the perturbation step size using the real-time P–V curve gradient, the algorithm effectively mitigates the trade-off between tracking speed and steady-state oscillations that occurs in conventional P\&O-based algorithms. A light-weight initialization routine enhances global maximum power point tracking under partial shading conditions. MPPT efficiency of 99.94\% with sub-millisecond convergence and low computational complexity is verified by comprehensive simulations for different irradiance profiles, temperatures, and converter topologies, and outperforms conventional P\&O and some intelligent MPPTs. Complementing the algorithm, this study provides a computational load evaluation and proposes a figure of merit for hardware-aware, scalable, and application-independent benchmarking. Future research directions are poised toward a mixed-signal integrated power management IC implementation featuring ADCs, DC–DC conversion, and on-chip MPPT logic, aiming to verify the algorithm performance under non-convex irradiance dynamics as a head-mounted optical receiver. These innovations will facilitate autonomous, self-powered microsystems in bio-integrated and wearable applications. Keeping the simplicity of the conventional P\&O algorithm with gradient-based adaptation, the proposed algorithm offers a lightweight solution for next-generation self-powered optical wireless power transfer microsystems, with a clear path toward efficient hardware integration.


\section*{Acknowledgment}



The authors thank Prof. Francesco Battaglia’s team for rodent locomotion data and Arash Akhoundi, PhD candidate at TU Delft, for technical support in algorithm development and signal processing.

\balance

\bstctlcite{IEEEexample:BSTcontrol}
\input{bare_jrnl_new_sample4.bbl}

\end{document}